\documentclass[a4paper,11pt]{article}
\usepackage{pos}

\title{The $N/D$ study on the singularity structure of $\pi N$ scattering amplitudes }

\author{Qu-Zhi Li\footnote{In collaboration with Han-Qing Zheng }}

\affiliation{Department of Physics and State Key Laboratory of Nuclear Physics and Technology,\\
  Peking University,
Beijing 100871, P. R. China}

\emailAdd{2001110075@stu.pku.edu.cn}

\abstract{The $N/D$ method is used to study the $S_{11}$ channel low energy $\pi N$ scattering amplitude. The input of left cuts are obtained from various phenomenological models. With the aid of the production representation, the total phase shifts can be decomposed into different contributions, and it further reveals that the existence of subthreshold resonance $N^*(890)$ doesn't depend on the details of the dynamical input. Additionally, it is found that there exist virtual states in partial waves, which are induced by the $u$ channel nucleon exchanges. These virtual states accumulate at the end point of the $u$ channel segment cut. The end point is hence the essential singularity of the full amplitude on the second sheet of complex $s$ plane.}

\FullConference{%
  10th International workshop on Chiral Dynamics - CD2021 \\
  15-19 November 2021 \\
  Online 
}

\begin{document}
\maketitle

\section{Introduction}
The $\pi N$ scatterings, as one of  the most fundamental processes in hadron physics, has been researched systemically for decades .    In  a series of recent publications \cite{WC43,WF14,WE}, however, it is suggested that there exists a sub-threshold $1/2^-$ nucleon resonance hidden in $S_{11}$ channel of $\pi N$ scatterings, with a pole mass $\sqrt s = (0.895\pm
0.081) -(0.164 \pm 0.023)i$ GeV. The result is obtained by using the production representation
(PKU representation) for partial wave amplitudes \cite{ZA733,ZA775,ZJ02,XA695,XB536}. The $N^*(890)$ pole may also be generated from a  K-matrix fit, though the
method suffers from the existence of spurious poles \cite{NC72}. Properties of $N^*(890)$ are also investigated, such as its coupling to $N\gamma$ and $N \pi$ \cite{MC45,CD11}. It is
found that its coupling to $N \pi$ is considerably larger than that of the $ N^*(1535)$, while its coupling
to $N\gamma$ is comparable to that of the $ N^*(1535)$. These results on couplings look reasonable and
are within expectations, hence providing further evidence on the existence of $N^*(890)$.

This talk   reviews  how to    directly find the  pole in the $S$ matrix element calculated from the N/D method, which comprises the Sec.II-IV.  The Sec.V focuses on the essential singularity of full $\pi N$ amplitude on the second sheet of complex $s$ plane.   
\section{A brief introduction to $N/D$ method}
The $N/D$ method is derived by Ref.~\cite{FP119} to restore the unitarity of the $\pi \pi$ partial wave amplitudes. For a single channel scattering, the partial wave amplitudes can be written as:
\begin{equation}
    T(s)=N(s)/D(s)\ .
\end{equation}
 where $D$ contains only the s-channel unitarity cut or the right hand cut $R$,   whereas $N$ only
contains the left hand cut ($l.h.c.$) or $L$. In single channel approximation, one has:
\begin{equation}
    \mathrm{Im}_R T(s)= \rho(s) |T(s)|^2\ , 
\end{equation}
where:
\begin{equation}
    \rho(s) = \dfrac{\sqrt{(s-s_L)(s-s_R)}}{s}\ ,
\end{equation}
with $s_L=(m_\pi-m_N)^2$, $s_R=(m_\pi+m_N)^2$. This leads to following relations:
\begin{align}\label{IMDN}
    &\mathrm{Im}_R D(s)=-\rho(s)N(s)\ ,\notag\\
   &  \mathrm{Im}_L N(s)= \mathrm{Im}_L[T(s)]D(s)\ .
\end{align}
Then one can write the dispersion relations:
\begin{align}\label{DRDN}
    & D(s) = 1 - \dfrac{(s-s_0)}{\pi}\int_R\dfrac{N(s^\prime)\rho(s^\prime)}{(s^\prime-s_0)(s^\prime-s)}ds^\prime\ ,\notag\\
    & N(s) = N(s_0) + \dfrac{(s-s_0)}{\pi}\int_L\dfrac{\mathrm{Im}_L[T(s^\prime)]D(s^\prime)}{(s^\prime-s_0)(s^\prime-s)}ds^\prime\ .
\end{align}
Noticing that when there appears circular cut in $T$ on $s$ plane,  We just need to make the following substitution: $\mathrm{Im}_L \to \dfrac{\mathrm{disc}_L}{2i}$ in Eq.~(\ref{IMDN}) and Eq.~(\ref{DRDN}).

To solve the Eq.~(\ref{DRDN}) one may substitute $D$ function into $N$ function and get a integral equation about $N$ function:
\begin{align}\label{NIEQ}
    N(s)= N(s_0) + \tilde{B}(s,s_0)+\dfrac{s-s_0}{\pi}\int_R\dfrac{\tilde{B}(s^\prime,s)\rho(s^\prime)N(s^\prime)}{(s^\prime-s_0)(s^\prime-s)}ds^\prime\ ,
\end{align}
with:
\begin{equation}\label{BEQ}
\tilde{B}(s^\prime,s)=\dfrac{s^\prime-s}{2i\pi}\int_L\dfrac{\mathrm{disc}_L[T(\tilde s)]}{(\tilde s-s)(\tilde s-s^\prime)}d\tilde s
\end{equation}
and use the inverse matrix method to obtain a numerical solution\footnote{In fact, we need a cutoff $\Lambda_R$ and we will set the $\Lambda_R^2=1.48 \mathrm{GeV^2}$ as well as $s_0=1\mathrm{GeV^2}$ at the following if  not specified. Introducing  $\Lambda_R$ means the $R$ in Eq.~(\ref{NIEQ}) presents the interval $(s_R,\Lambda_R^2)$.}. 

After getting a numerical solution, the next step is to  search for the pole on the second sheet, hence analytical continuation to the complex plane is necessary: 
\begin{equation}
    D^{II}(s) = D^I(s) +2i\rho(s)N(s),\qquad N^{II}(s)=N^I(s)\ . 
\end{equation}
\section{The PKU representation}
Elastic partial wave S matrix elements satisfy a production representation
like follows:
\begin{equation}\label{pkue}
    S = \prod_i S_i\times e^{2 i \rho(s) f(s)}\ ,
\end{equation}
Detailed discussions on how to obtain Eq.~(\ref{pkue}) can be found in Refs.~\cite{ZA733,ZA775}, which is fully consistent with unitarity, analyticity and crossing sysmmetry \cite{Guo20h,Guo20f}.  For $\pi \pi$ system or $\pi K$ system, the "spectral" function $f(s)$  reads :
\begin{equation}\label{Cf}
    f(s) = \ln\left(\dfrac{S}{\prod_i S_i}\right)/(2i\rho(s))\ .
\end{equation}
In $\pi N$ system, however, the situation is different.  There exists a point $s_c\simeq m_N^2-m_\pi^4/(2m_N^2)$  in the segment cut $(c_L,c_R)$.  The real part of the argument of  logarithmic function  is  negative near the point,  but the imaginary part will change sign when s crosses $s_c$, which leads to that  $f(s)$  has to develop a discontinuity and hence a branch
cut emerges crossing $s_c$. So we redefine $f(s)$ as the follwing:
\begin{equation}\label{CfR}
    f(s)=\ln\left(-\dfrac{S}{\prod_i S_i}\right)/(2i\rho(s))-\dfrac{\pi}{2\bar \rho(s)}\ .
\end{equation}
where the function $\bar \rho(s) $ is the `deformed' $\rho(s) $ with its cut $\in [s_L,s_R]$, while the cut of the
latter is defined on $(-\infty,s_L) \cup (s_R,\infty)$. Notice that $\bar\rho(s)$ and $\rho(s)$ are identical in the physical
region. In fact, for the  same process, Eq.~(\ref{Cf}) is equal to  Eq.~(\ref{CfR})  in the physical region, but their left cuts are different. The definition of Eq.~(\ref{CfR}) simplifies the $f(s)$ cut structure, so using Eq.~(\ref{CfR}) can simplify PKU decomposition. At last, we emphasize that the $S$ matrix only possesses  physical cut and primordial left cuts .
\section{Numerical calculations}
In Ref.~\cite{LC46}, The authors do several  calculations with various   $\mathrm{Im}_LT(s)$ as the input of $N/D$ method. Different input lead the unitary amplitudes $T(s)$ possessing different $l.h.c.$, but in every calculation, one   can find a pole on the second sheet. The details can be seen in the following.    
\subsection{Toy model calculations}
Firstly, let's see a solvable example. Assuming $\mathrm{Im}_LT(s)$ is simulated by a set of Dirac $\delta$ functions, or equivalently:
\begin{equation}\label{NPI}
    N(s) = \sum_i\dfrac{\gamma_i}{s-s_i}\ ,
\end{equation}
which is to be used in the first equation of Eq.~~(\ref{DRDN}). We (arbitrarily) choose $ Case  I$: one pole at $s_1$ = 0, and $Case II$: one pole at $s_1 =-m_N^2$ , and fit to
the "data" obtained from the solutions of Roy Steiner equations\cite{MPR625} by tuning the parameter $\gamma_1$, and search for poles on the $s$-plane. Both cases give a good fit to the data, and a sub-threshold
pole emerges in each case with a location listed in Table~\ref{poleposition}.
	\begin{table}[h]
		\centering
		\begin{tabular}[width=12cm]{|c|c|c|}
			\hline
			& $Case$ I & $Case$ II \\
			\hline
			$s_1$ &  0 & $-m_N^2$ \\
			\hline
			$\gamma_1$ (GeV$^2$) &  0.79 & 1.34 \\
			
			\hline
			$\sqrt{s_{pole}}$(GeV) &  0.810 - 0.125i & 0.788 - 0.185i \\      \hline
		\end{tabular}
		\caption{Subthreshold pole locations using input Eq.~~(\ref{NPI}).}\label{poleposition}.
	\end{table}

The  phase shift and its PKU decomposition  are plotted in Ref.~\cite{LC46}, and it tells us the background cut ($l.h.c.$) contribution to the phase shift is concave and negative while the subthreshold
resonance pole provides a positive and convex phase shift above threshold to counterbalance the
former contribution, and the sum of the two reproduces the steadily rising phase shift data. 
\subsection{N/D calculations using pure $\chi P T$ input}
A very natural idea is that using the $\mathrm{Im}_L T$ obtained from $\chi PT$  as the input of $N/D$.  As mentioned in Ref.~\cite{LC46}, however, the $\mathrm{Im}_L T$  encounters the
problem:
\begin{equation}
    \mathrm{Im}_L T[\mathcal{O}(p^n)](s\to 0)\sim Con.\times s^{-n-1/2}\ , \quad n>1\ ,
\end{equation}
where  $n$ is the chiral order. It's in contradiction to Froissart bound, but according to the Regge model, when $s\to 0$, the  partial wave amplitude should behave as  :
\begin{equation}
    T\sim s^{-\alpha(0)}\ ,
\end{equation}
with $\alpha(0) \simeq0.19$. 
Though    $\mathrm{Im}_LT$ of $\chi$PT encounters such a problem, there still exist two ways to avoid this singularity when use it   as the input of $N/D$ method. 

One way is like the Ref.~\cite{LC46}, and  let the auxiliary function $\tilde B(s^\prime,s)$ in eqs.(\ref{NIEQ}) and (\ref{BEQ}) formally be written  as $\tilde B(s^\prime,s)=T_L(s^\prime)- T_L(s)$, where $T_L$ is taken as the $\mathcal{O}(p^1)$ partial wave
amplitude plus the $\mathcal{O}(p^2)$ part of $T^J_{+-}\footnote{The definition can be found in Ref.~\cite{LC46}}.$.

At $\mathcal{O}(p^2)$ level there are four low energy constants (LECs) $c_i$ with $i=1,\dots,4$. A good fit is obtained with $c_1=-0.40,~ c_2 = 3.50,~ c_3 = -3.90,~ c_4 = 2.17,~ N(s_0) = 0.47$. Of cause, these LECs satisfy the positivity constraints\cite{CE74}. The pole locates at :
\begin{equation}
    \sqrt{s} = (1.01-0.19i) \mathrm{ GeV}\ .
\end{equation}

Another ways is to introduce intermediate function $\bar T(s)$ and the partial wava amplitude $T(s) = \bar T(s)/s^2$, where $\bar T(s) $ can be written as :
\begin{equation}
    \bar T(s) = \dfrac{N(s)}{D(s)}\ ,
\end{equation}
According to  unitary, the $N(s)$ and $D(s)$ functions should satisfy at this time:
\begin{align}
    & \mathrm{Im}_L[N(s)]=D(s)\mathrm{Im}_L[\bar T(s)]=D(s) s^2\mathrm{Im}_L[T(s)] \ ,\notag\\
    & \mathrm{Im}_R[D(s)]=N(s)\mathrm{Im}_R[\bar T^{-1}(s)]=-\rho(s)\dfrac{N(s)}{s^2}\ .
\end{align}
Taking $N(s)$ into $D(s)$ function after writing dispersion relations for them, one can get an integral function about $D(s)$:
\begin{equation}\label{DIEQ}
    D_r(s) = \dfrac{1}{s-s_0}-a_0F(s,s_0)+\dfrac{1}{\pi}\int_LF(s,s^\prime)s^{\prime2} \mathrm{Im}_L[T(s^\prime)]D_r(s^\prime)ds^\prime
\end{equation}
with:
\begin{equation}\label{FF}
    D_r=\dfrac{D(s)}{s-s_0},\quad F(s,s_0)=\dfrac{1}{\pi}\int_R\dfrac{\rho(s^\prime)ds^\prime}{s^{\prime2}(s^\prime-s)(s^\prime-s_0)}\ .
\end{equation}
Now we need a cutoff \footnote{For example, Eq.~\ref{ac1} is obtained by fixing $\Lambda^2_R=5\mathrm{GeV^2}$. } in Eq.~(\ref{FF}) and an extra cutoff $\Lambda_L(\Lambda_L^2=1 GeV^2)$ in Eq.~(\ref{DIEQ}), which meas the $L$ and $R$ denote the interval$(-\Lambda_L^2,s_L)$ and $(s_R,\Lambda_R^2)$, respectively. Then  Eq.~(\ref{DIEQ}) is solved numerically on the interval $(-\Lambda_L^2,s_L)$. Again we fit the data  using the subtraction constant $a_0$ and
the LECs in the $\mathcal{O}(p^2)$ $\chi PT$ as fit parameters and get:
\begin{equation}\label{ac1}
    a_0 = 0.88, c_1 = -0.40, c_2 = 3.10, c_3 = -3.50, c_4 = 4.00\ .
\end{equation}
One pole  locates at :
\begin{equation}
    \sqrt{s}=(0.91-0.2i) \mathrm{GeV}\ ,
\end{equation}
even when we push the cutoff $\Lambda_R^2$
 to infinity, $ N^*(890)$ still exists:
 \begin{equation}
     \sqrt{s} = (0.93-0.21i)\mathrm{GeV}\ .
 \end{equation}
\subsection{N/D calculation using phenomenological models}
The  ill singularities at $s = 0$ in partial wave chiral amplitudes, as mentioned above, come at least partly from integrating out heavy degrees of freedom, such as: the $\rho$ exchanges and $\dfrac{1}{2}^\pm$  baryon exchange etc. Before integrating out heavy degrees of freedom, all these resonance exchange amplitudes contain singularity of $s^{-1/2}$ type at most when $s=0$. This reveals the fact that partial wave projections and chiral expansions do
not commute. 

On the other hand, when we make an asymptotic expansion of heavy degrees of freedom exchange contributions to $T (S_{11})$ in the vicinity of $s=0$, we find that the first two most singular terms are of type:
\begin{equation}
    \dfrac{a+bs}{\sqrt{s}}\ ,
\end{equation}
which pushes us to  use  $\mathcal{O}(p^1) \chi PT $ results plus the
$\rho$ meson exchange term and a polynomial as the input of $N/D$ method: 
\begin{equation}
    \mathrm{disc} T(s) = \mathrm{disc} T^{(1)} + \mathrm{disc} T^{(\rho)}+ \mathrm{disc}\left[\dfrac{a+bs}{\sqrt{s}}\right]\ .
\end{equation}
where the $\rho$ meson exchange term can be in charge of part the contributions of  circular cut. The fit gives $N(s_0)=0.61,\ a=-7.88GeV\ ,b=-8.00GeV^{-1}$. Finally, one second sheet pole is found located at:
\begin{equation}
    \sqrt{s} = (0.90 - 0.20 i)\mathrm{GeV}\ .
\end{equation}

At the end of this section, it should be emphasized that the PKU decomposition is done for every calculation and the results are similar:  the background cut contribution to the phase shift is  negative while the  pole gives a positive and phase shift, and the sum of the two reproduces the  phase shift data. These calculations also tell us the existence of $N^*(890)$ dose not  depend on the model details.

\section{Essential Singularities of $\pi N$ Scattering
amplitudes}
It is pointed out that in the $L_{2I,2J}=S_{11}$ channel partial wave $\pi N$ scattering amplitude, there exist two virtual poles, according to the  Ref.~\cite{LC46} . In fact, the existence of virtual poles are quite universal, but
different channels behaves rather differently. For simplicity,  definitions of notation  are ignored there, and one can find them in Ref.~\cite{LA21}.

The relations between  partial wave helicity amplitudes and invariant amplitudes function $A^I$ as well as  $B^I$(I:isospin ) are expressed as following:
\begin{equation}\label{PW}
	\begin{split}
	&T^{I,J}_{++}(s)=\dfrac{1}{64\pi}[2m_NA_C^I(s)+(s-m_\pi^2-m_N^2)B_C^I(s)]\\
	&T^{I,J}_{+-}(s)=-\dfrac{1}{64\pi\sqrt{s}}[(s-m_\pi^2+m_N^2)A^I_S(s)+m_N(s+m_\pi^2-m_N^2)B^I_S(s)]\ \,
	\end{split}
\end{equation}  
with
\[F_C^I(s) = \int_{-1}^{1}dz_sF^I(s,t)[P_{J+1/2}(z_s)+P_{J-1/2}(z_s)]\ ;\]
\[F_S^I(s) = \int_{-1}^{1}dz_sF^I(s,t)[P_{J+1/2}(z_s)-P_{J-1/2}(z_s)]\ ,\]
where $F$ stands for $A$ or $B$. Meanwhile, the scalar function $A^I$ and $B^I$ can be calculated by ChPT. Here, we presented the results of $\mathcal{O}(p^1)$
\begin{equation}\label{ABF}
	\begin{split}
	& A^{1/2}(s,u)=A^{3/2}(s,u)=\dfrac{m_Ng^2}{F^2}\ ,\\
	& B^{1/2}(s,u)=\dfrac{1-g^2}{F^2}-\dfrac{3m_N^2g^2}{F^2(s-m_N^2)}-\dfrac{m_N^2g^2}{F^2}\dfrac{1}{u-m_N^2}\ ,\\
	& B^{3/2}(s,u)=\dfrac{g^2-1}{F^2}+\dfrac{2m_N^2g^2}{F^2}\dfrac{1}{u-m_N^2}\ ,
	\end{split}
\end{equation}
where $F$ and $g$ denote the pion decay constant and the axial coupling constant, respectively. Here we however only need to concern the $ 1/(u-m^2_N
) $ term, which coefficient is immune of
any chiral corrections, and its sign determines the existence of the zeros (poles) of the $S$-matrix.
Taking Eq.~\ref{ABF}   into   Eq.~\ref{PW}, and using the Neumann equation:
\begin{equation}
	Q_l(x)=\dfrac{1}{2}\int_{-1}^{1}\dfrac{P_l(y)}{x-y}dy\ ,
\end{equation}
where $Q_l$ is called the second Legendre function, one can get 
\begin{equation}
\begin{split}
&	B^{1/2,J}_{C,S} =  -\dfrac{m_N^2g^2}{16\pi F^2 s\rho(s)^2}[Q_{J+1/2}(y(s))\pm Q_{J-1/2}(y(s))]+\dots\ ,\\
&   B^{3/2,J}_{C,S} = \dfrac{m_N^2g^2}{8\pi F^2 s\rho(s)^2}[Q_{J+1/2}(y(s))\pm Q_{J-1/2}(y(s))]+\dots\ .
\end{split}
\end{equation}
in which the neglected terms $\dots$ denote the parts regular at $c_L=(m_N^2-m_\pi^2)^2/m_N^2$ and $c_R=m_N^2+2m_\pi^2$ (see discussions below),
which receive chiral corrections. The definition of $y(s)$ is
\begin{equation}
	y(s)=\dfrac{2m_\pi^2s-s^2+s_Ls_R}{(s-s_L)(s-s_R)}\ .
\end{equation}
Using the explicit expressions of $Q_l(s)$
\begin{equation}
	\begin{split}
&	Q_0(x)=\dfrac{1}{2}\ln\dfrac{x+1}{x-1}\ ,\\
& Q_l(x)=\dfrac{1}{2}P_l(x)\ln\dfrac{x+1}{x-1}-O_{l-1}(x)\ ,
\end{split}
\end{equation}
with 
\begin{equation}\label{OD}
	O_{l-1}(x)=\sum_{r=0}^{\left[\dfrac{l-1}{2}\right]}\dfrac{(2l-4r-1)}{(2r+1)(l-r)}P_{l-2r-1}(x)\ .
\end{equation}
One can find that the function $B^{I,J}_{C,S}(s)$ has the term 
\begin{equation}
	\ln\dfrac{y(s)+1}{y(s)-1}=\ln\dfrac{m_N^2}{s}+\ln\dfrac{s-c_L}{s-c_R} \ .
\end{equation}
This term leads to that there exist three branch points at $s=0,c_L,c_R$ in function $B^{I,J}_{C,S}$, and then in parity eigenstates amplitudes $T^{I,J}_{\pm}$ . When $s$ tends to $c_R$ , it's easy to find that 
\begin{equation}
\begin{split}
&	s\to c_R,\quad T^{1/2,J}_{\pm}\to \frac{g^2 m_N^2 (m_N^2+2 m_\pi^2)} {16\pi  F^2 (4 m_N^2-m_\pi^2)}\ln \frac{c_R-c_L}{s-c_R}\to \infty\ , \\
&	s\to c_R,\quad T^{3/2,J}_{\pm}\to -\frac{g^2 m_N^2 (m_N^2+2 m_\pi^2)} {8\pi  F^2 (4 m_N^2-m_\pi^2)}\ln \frac{c_R-c_L}{s-c_R}\to -\infty\ .	
\end{split}
\end{equation}
The sign is independent of the angular momentum $J$.  Things will be different, however, when $s$ tends to $c_L$. It turns out that $ T^{1/2,J}_{\pm} $ tends to $ \mp(-1)^{J+1/2}\infty$, and  $ T^{3/2,J}_{\pm}$ tends to $ \pm (-1)^{J+1/2}\infty$. Taken the definition  of  $S$-matrix  and the fact that $2i\rho(s)$ is negative at $s=c_R,c_L$ into consideration, conclusions can be draw that:
\begin{itemize}
	\item [1)]  In the region $s\in(c_R,s_R)$, the S-matrix $S^{1/2,J}_{\pm}(s)$ must exist a zero point, which corresponds to a virtual state.
	\item [2)] In the region $s\in(s_L,c_L)$, the S-matrix $S^{1/2,J}_{+}(s)$ and $S^{3/2,J}_{-}(s)$   exist  zero points for  $J=1/2, 5/2, 9/2,...$ ; the S-matrix $S^{1/2,J}_{-}(s)$ and $S^{3/2,J}_{+}(s)$   exist  zero points for  $J=3/2, 7/2, 11/2,...$.
\end{itemize}

The following discussions dedicate to determining the location of those zeros. Let us focus
on I = 1/2 for the moment.
 The  explicit expressions of S-matrix reads
\begin{equation}
		S^{1/2,J}_\pm(s) = \mathcal A^{1/2,J}_{\pm}(s)+\mathcal B^{1/2,J}_{\pm}(s)\ln\dfrac{s-c_L}{s-c_R}\ ,\ \quad J\geq \dfrac{3}{2}\  ,
\end{equation}
with
\begin{align}\label{A1J}
\mathcal A^{1/2,J}_{\pm}(s)=1-\dfrac{im_N^2g^2}{4\pi F^2 s\rho(s)}\left\{(W\pm m_N)(E_N\mp m_N)[\dfrac{P_{J+1/2}(y)}{2}\ln\dfrac{m_N^2}{s}-O_{J-1/2}(y)]+\notag\right.
\\
\phantom{=\;\;}
\left.(W\mp m_N)(E_N\pm m_N)[\dfrac{P_{J-1/2}(y)}{2}\ln\dfrac{m_N^2}{s}-O_{J-3/2}(y)]
\right\}\ ,
\end{align}
where $W\equiv \sqrt{s}$ denotes the  center-of-mass frame energy and $E_N=\dfrac{s+m_N^2-m_\pi^2}{2\sqrt{s}}$ is the nucleon energy. It's worth  stressing that the function $\mathcal B^{1/2,J}_{\pm}(s)$  is immune of chiral perturbation corrections. 

Let $v^J_{R\pm}\in (c_R,s_R)$ is the zero point of $S^{1/2,J}_{\pm}(s)$,which gives 

\begin{equation}\label{VR}
S^{1/2,J}_{\pm}(v^J_{R\pm})=\mathcal A^{1/2,J}_{\pm}(v^J_{R\pm})+\mathcal B^{1/2,J}_{\pm}(v^J_{R\pm})\ln\dfrac{v^J_{R\pm}-c_L}{v^J_{R\pm}-c_R}=0\ .
\end{equation}  

Supposing $v^J_{R\pm}\approx c_R$ when $J$ greater than some $J_N$ , the solution of Eq.~\ref{VR}  reads
\begin{equation}\label{VRS}
		v^J_{R\pm}=c_R+(c_R-c_L)e^{\mathcal A^{1/2,J}_{\pm}({c_R})/\mathcal B^{1/2,J}_{\pm}({c_R})}\ .
\end{equation}
Using  $P_l[y(c_R)] =P_l(1)=1$, one can find that $\mathcal B^{1/2,J}_{\pm}({c_R})$ is negative and independent of $J$. The function $\mathcal A^{1/2,J}_{\pm}({c_R}) $  increases as $J$ increases, since $ O_{J+1/2}(1)-O_{J-1/2}(1) $ is greater than zero according to  Eq.~\ref{OD}. Further, when $J$ tends $\infty$, the $\mathcal A^{1/2,J}_{\pm}({c_R}) $ also goes to the infinity, which can be seen through 
\begin{equation}
		\begin{split}
	\lim_{n\to\infty} O_{n}(1)&=\lim_{k\to\infty}O_{2k}(1)=\lim_{k\to\infty}\sum_{r=0}^{k}\dfrac{(4k-4r+1)}{(2r+1)(2k+1-r)}\\
	&=\lim_{k\to\infty}\sum_{r=0}^{k}\left(\dfrac{2}{2r+1}-\dfrac{1}{2k-r+1}\right)>\lim_{k\to\infty}\sum_{r=0}^{k}\left(\dfrac{2}{2r+1}-\dfrac{1}{k+1}\right)
	\\
	&=\lim_{k\to\infty}\sum_{r=0}^{k}\dfrac{2}{2r+1}-\lim_{k\to\infty}\sum_{r=0}^{k}\dfrac{1}{k+1}\to\infty
	\ .
	\end{split}
\end{equation} 
It means the zero point, $v_{R\pm}$, tends $c_R$ with the $J$ increasing  so that $v_{R\pm}=c_R$ when $J\to\infty$ according to Eq.~\ref{VRS}. The zero point of partial wave $S$-matrix on the first sheet will become the pole of partial wave amplitude on the second sheet according to the analytic continuation:
\begin{equation}
    T^{\mathrm{II}}(s) = \dfrac{T(s)}{S(s)}\ .
\end{equation}
Moreover, full amplitudes on the second sheet can be written as :
\begin{equation}
	T_{+\pm}^{\mathrm{II}}(s,t) = 8\pi\sum_{J=1/2}(2J+1)\left[\dfrac{	T^J_+(s)}{	S^J_+(s)} \pm \dfrac{	T^J_-(s)}{	S^J_-(s)}\right]d^J_{1/2,\pm 1/2}(\cos\theta)\ .
\end{equation}
Now one can find that the full amplitude $T^{\mathrm{II}}_{+\pm}$ possesses infinite poles  on a finite interval, so these poles will accumulate and $c_R$ is the accumulation point  on the second sheet. Similarly analyses can
be made in the situation in the line segment $(s_L, c_L)$. Further, one finds that both $v^{1/2,J}_{L+}$ and $v^{1/2,J}_{L-}$
approaches $c_L$ when $J\to\infty$. Therefore, following the steps of Ref.~\cite{MF19}, we prove that $s = c_L, c_R$ are two essential singularities of $T^{1/2}(s,t)$, on the second sheet of complex $s$-plane. Unlike  Ref.~\cite{MF19}, where the proof is fully non-perturbative, our proof given here is valid
to all orders of perturbation chiral expansions. As for $I=3/2$, one can prove that only $c_L$ is an essential
singularity of $T^{3/2}(s,t)$ on the second sheet of s plane.

This work is supported in part by National Nature Science Foundations of China (NSFC) under Contracts No. 11975028 and
No. 10925522.

\end{document}